\documentclass[11pt,a4paper]{article}
\usepackage{latexsym,amsmath,amssymb}
\usepackage{amsthm}
\usepackage{bm}
\usepackage{cite}
\usepackage{microtype}
\usepackage{t1enc}
\usepackage{xspace}
\usepackage{pdfpages}
\usepackage{color}
\usepackage[breaklinks]{hyperref}
\usepackage{ifpdf}
\usepackage{graphicx}
\usepackage{verbatim}
\usepackage{tikz}
\usepackage{pgf}
\usepackage{pstricks}
\usepackage{mathrsfs}
\usepackage{paralist}
\usepackage{amsfonts}
\usepackage{dsfont}
\usepackage{wasysym}
\usepackage{esvect}
\usepackage{tabularx}
\usepackage{url}
\usepackage{subfig}
\usepackage{listings}
\usepackage{epic}

\newcommand{\set}[2]{\{\, {#1}\mid{#2}\,\}}

\renewcommand{\emptyset}{\varnothing}
\renewcommand{\epsilon}{\varepsilon}

\newcommand{\clockconstraints}[1]{\Phi(#1)}
\newcommand{\forwardclock}[1]{\overrightarrow{x_{#1}}}
\newcommand{\backwardclock}[1]{\overleftarrow{x_{#1}}}
\newcommand{\forwardstackclock}{\overrightarrow{x_{\mathrm{pop}}}}
\newcommand{\backwardstackclock}{\overleftarrow{x_{\mathrm{push}}}}

\newtheorem{lemma}{Lemma}
\newtheorem{claim}{Claim}
\newtheorem{theorem}{Theorem}
\newtheorem{definition}{Definition}
\newtheorem{example}{Example}

\begin{document}

\sloppy

\title{On the determinization of event-clock input-driven pushdown automata\thanks{%
	Supported by the Russian Foundation for Basic Research
	under grant 20-51-50001.}
}

\author{Mizuhito Ogawa\thanks{%
	Japan Advanced Institute of Science and Technology, Japan,
	\texttt{mizuhito@jaist.ac.jp}.}
	\and
	Alexander Okhotin\thanks{%
	Department of Mathematics and Computer Science,
	St.~Petersburg State University, 7/9 Universitetskaya nab., Saint Petersburg 199034, Russia,
	\texttt{alexander.okhotin@spbu.ru}.}
}

\maketitle

\begin{abstract}
Input-driven pushdown automata (also known as visibly pushdown automata
and as nested word automata) are a subclass of deterministic pushdown
automata and a superclass of the parenthesis languages.
Nguyen and Ogawa
(\href{https://doi.org/10.1007/978-3-540-95891-8_50}{``Event-clock visibly pushdown automata''}, SOFSEM 2009)
defined a timed extension of these automata under the event-clock model,
and showed that this model can be determinized using the method of region construction.
This paper defines
a further extension of this model with the event clock on the call-return operations,
and proposes a new, direct determinization procedure for these automata:
an $n$-state nondeterministic automaton
with $k$ different clock constraints
is transformed to a deterministic automaton with $2^{n^2}$ states,
$2^{n^2+k}$ stack symbols and the same clock constraints as in the original automaton.
The construction is shown to be asymptotically optimal
with respect to both the number of states and the number of stack symbols. \\

\end{abstract}

\section{Introduction}

Timed automata (TA), introduced by Alur and Dill~\cite{AlurDill},
are finite automata operating in real time.
These automata enjoy decidable emptiness problem
(equivalently, the state reachability problem) 
and is implemeneted as UPPAAL~\footnote{\tt http://www.uppaal.org/} for safety checking. 
The decidability of emptiness holds under various extensions with the pushdown stack,
such as
{\em Dense-Timed Pushdown Automata} (DTPDA) of Abdulla et al.~\cite{AbdullaAtigStenman}
with {\em ages} (representing local clocks), 
which are further analyzed by Clemente and Lasota~\cite{ClementeLasota}. 

Although the emptiness problem for timed automata is decidable,
timed automata are not closed under complementation,
and their nondeterministic case cannot generally be determinized.
Their inclusion problem is decidable only in the case of a single clock~\cite{OuaknineWorrell},
and for two or more clocks it becomes undecidable~\cite{AlurDill}.

As an alternative timed device,
the class of {\em event-clock automata} (ECA)
was introduced by Alur et al.~\cite{AlurFixHenzinger}
and further studied by Geeraerts et al.~\cite{GeeraertsRaskinSznajder}:
this class allows determinization and complementation,
and hence it enjoys decidable inclusion problem.
An ECA is defined with a ``prophecy clock'' and a ``history clock'' bound to each input symbol. 
The history clock $\backwardclock{a}$ associated to an input symbol $a$
is always reset when $a$ is read, 
and the prophecy clock $\forwardclock{a}$ predicts the next occurrence of $a$. 

In general, when a pushdown stack is introduced,
this often destroys the decidablity of the inclusion problem,
since asynchronous behavior of two pushdown stacks
disrupts a direct product of two devices.
Even starting from finite automata,
adding the pushdown stack makes the inclusion undecidable.

To remedy this, a constraint on the synchronous behaviour of pushdown stacks
is imposed upon the model.
The resulting {\em input-driven pushdown automata}~\cite{Mehlhorn,vonBraunmuehl_Verbeek} (IDPDA),
also known as {\em visibly pushdown automata}~\cite{AlurMadhusudan}
and as {\em nested word automata}~\cite{AlurMadhusudan2},
are defined over an alphabet split into three parts:
\emph{left brackets} $\Sigma_{+1}$, on which the automaton must push one stack symbol,
\emph{right brackets} $\Sigma_{-1}$, on which the automaton must pop one stack symbol,
and \emph{neutral symbols} $\Sigma_{0}$, on which the automaton ignores the stack.
Unlike the standard pushdown automata, IDPDA are closed under all Boolean operations,
and they can be determinized.
There is a Myhill--Nerode-like characterization for these automata~\cite{AlurKumarMadhusudanViswanathan}.
Major contributions of Alur and Madhusudan~\cite{AlurMadhusudan,AlurMadhusudan2}
include a lower bound on the number of states needed to determinize these automata,
which started a line of research
on the succinctness of description for this model~\cite{idpda_sigact_survey},
and a B\"uchi-like extension for $\omega$-words~\cite{LoedingMadhusudanSerre,OkhotinSelivanov}.

Combining the ideas of input-driven pushdown and event-clock automata,
{\em event-clock visibly pushdown automata}
were proposed by Nguyen and Ogawa~\cite{NguyenOgawa},
followed and extended by Bhave et al.~\cite{BhaveDaveKrishnaPhawadeTrivedi}
and Bozzelli et al.~\cite{BozzelliMuranoPeron}.
This paper revisits this model,
with the aim to investigate the determinization and the emptiness problem,
which leads the decidability of the inclusion problem.
We further extend the model by introducing special event clocks
recording the duration of the call/return relation.
The resulting model is called {\em event-clock input-driven pushdown automata} (ECIDPDA).
We observe that the Boolean operations and the determinization
work in the presence of event clocks on the call/return relation.
The determinization is \emph{direct},
in the sense that it \emph{does not rely on the classical
discretization} or ``untime translation'' method,
which allows the bisimulation of timed transitions to be maintained,
and \emph{is not based on the region construction},
which handles the extension by the age of a stack symbol 
in Bhave et al.~\cite{BhaveDaveKrishnaPhawadeTrivedi}.

As per the proposed construction,
presented in Section~\ref{section_ECIDPDA_determinization},
any given $n$-state nondeterministic automaton
with $k$ different clock constraints and with any number of stack symbols
is transformed to a deterministic automaton with $2^{n^2}$ states,
$2^{n^2+k}$ stack symbols and the same clock constraints as in the original automaton.
Furthermore, in Section~\ref{section_ECIDPDA_determinization_lower_bound},
this construction is shown to be asymptotically optimal
both with respect to the number of states
and with respect to the number of stack symbols.

\section{Definitions}

Event-clock automata operate on \emph{timed strings} over an alphabet $\Sigma$,
that is, sequences of the form $w=(a_1, t_1) \ldots (a_n, t_n)$,
where $a_1 \ldots a_n \in \Sigma^*$ is a string,
and $t_1 < \ldots < t_n$ are real numbers indicating the time of the symbols' appearance.

For input-driven pushdown automata, the alphabet $\Sigma$ is split into three disjoint classes:
$\Sigma=\Sigma_{+1} \cup \Sigma_{-1} \cup \Sigma_0$,
where symbols in $\Sigma_{+1}$ are called \emph{left brackets},
symbols in $\Sigma_{-1}$ are \emph{right brackets},
and $\Sigma_0$ contains \emph{neutral symbols}.
An input-driven automaton always pushes one stack symbol upon reading a left bracket,
pops one stack symbol upon reading a right bracket,
and does not access the stack on neutral symbols.
Typically, a string over such an alphabet is assumed to be \emph{well-nested}
with respect to its left and right brackets,
but the most general definition of input-driven automata also allows ill-nested inputs.

The proposed \emph{event-clock input-driven pushdown automata} (ECIDPDA)
operate on timed strings
over an alphabet $\Sigma=\Sigma_{+1} \cup \Sigma_{-1} \cup \Sigma_0$.
These automata operate like input-driven automata,
and additionally can evaluate certain constraints
upon reading each input symbol.
These constraints refer to the following \emph{clocks}:
\begin{itemize}
\item
a symbol history clock $\backwardclock{a}$, with $a \in \Sigma$,
provides the time elapsed since the symbol $a$ was last encountered;
\item
a symbol prediction clock $\forwardclock{a}$, with $a \in \Sigma$,
foretells the time remaining until the symbol $a$ will be encountered next time;
\item
a stack history clock $\backwardstackclock$,
defined on a right bracket,
evaluates to the time elapsed since the matching left bracket;
\item
a stack prediction clock $\forwardstackclock$,
defined on a left bracket,
foretells the time remaining until the matching right bracket.
\end{itemize}
These values are formally defined as follows.

\begin{definition}\label{definition_clock}
Let $\Sigma = \Sigma_{+1} \cup \Sigma_{-1} \cup \Sigma_0$ be an alphabet.
The set of clocks over $\Sigma$ is
$\mathcal{C}(\Sigma) = \set{\backwardclock{a}}{a \in \Sigma}
 \cup \set{\forwardclock{a}}{a \in \Sigma}
 \cup \{\backwardstackclock, \forwardstackclock\}$.
Then, the value of a clock $C \in \mathcal{C}(\Sigma)$
on a timed string $w=(a_1, t_1) \ldots (a_n, t_n)$
at position $i \in \{1, \ldots, n\}$ is defined as follows.
\begin{itemize} 
\item The value of a symbol history clock $\backwardclock{a}$ on $w$ at $i$
is $t_i-t_j$, where $j \in \{1, \ldots, i-1\}$
is the greatest number with $a_j=a$.
If no such $j$ exists, the value of $\backwardclock{a}$ is undefined.

\item The value of a symbol prediction clock $\forwardclock{a}$ on $w$ at $i$
is $t_j-t_i$, where $j \in \{i+1, \ldots, n\}$
is the least number with $a_j=a$.
If no such $j$ exists, the value of $\forwardclock{a}$ is undefined.

\item The value of a stack history clock $\backwardstackclock$ on $w$ at $i$
is defined only if $a_i$ is a right bracket $a_i \in \Sigma_{-1}$,
and this bracket has a matching left bracket $a_j \in \Sigma_{+1}$
at a position $j<i$.
In this case, the value of $\backwardstackclock$ on $w$ at position $i$ is $t_i-t_j$;
otherwise it is undefined.

\item The value of a stack prediction clock $\forwardstackclock$ on $w$ at $i$
is defined only if $a_i$ is a left bracket $a_i \in \Sigma_{+1}$,
and this bracket has a matching left bracket $a_j \in \Sigma_{-1}$
at a position $j>i$.
In this case, the value of $\forwardstackclock$ on $w$ at position $i$ is $t_j-t_i$;
otherwise it is undefined.
\end{itemize}
\end{definition}

The original model by Nguyen and Ogawa~\cite{NguyenOgawa}
used only symbol history clocks $\backwardclock{a}$
and symbol prediction clocks $\forwardclock{a}$.
Stack history clocks $\backwardstackclock$
were first introduced by Bhave et al.~\cite{BhaveDaveKrishnaPhawadeTrivedi},
who called them \emph{the age of stack symbols}.
As compared to the definition of Bhave et al.~\cite{BhaveDaveKrishnaPhawadeTrivedi},
another clock type,
the stack prediction clock $\forwardstackclock$,
has been added to the model:
it is symmetric to the stack history clock $\backwardstackclock$.

A clock constraint is a logical formula that restricts the values of clocks at the current position:
clocks values can be compared to constants,
and any Boolean combinations of such conditions can be expressed.

\begin{definition}\label{definition_clock_constraint}
Let $\Sigma=\Sigma_{+1} \cup \Sigma_{-1} \cup \Sigma_0$ be an alphabet.
The set of clock constraints over $\Sigma$,
denoted by $\clockconstraints{\Sigma}$,
consists of the following formulae.
\begin{itemize}
\item
	For every clock $C \in \mathcal{C}(\Sigma)$
	and for every non-negative constant $\tau \in \mathbb{R}$,
	the following are atomic clock constraints:
	$C \leqslant \tau$;
	$C \geqslant \tau$.
\item
	If $\varphi$ and $\psi$ are clock constraints,
	then so are $(\varphi \lor \psi)$ and $(\varphi \land \psi)$.
\item
	If $\varphi$ is a clock constraint,
	then so is $\lnot \varphi$.
\end{itemize}

Let $w=(a_1, t_1) \ldots (a_n, t_n)$ be a timed string,
let $i \in \{1, \ldots, n\}$ be a position therein.
Each clock constraint can be either true or false on $w$ at position $i$,
which is defined inductively on its structure.
\begin{itemize}
\item
	A clock constraint $C \leqslant \tau$ is true
	if the value of $C$ on $w$ at position $i$
	is defined and is at most $\tau$.
\item
	A clock constraint $C \geqslant \tau$ is true
	if the value of $C$ on $w$ at position $i$
	is defined and is at least $\tau$.
\item
	$(\varphi \lor \psi)$ is true on $w$ at $i$,
	if so is $\varphi$ or $\psi$;
\item
	$(\varphi \land \psi)$ is true on $w$ at $i$,
	if so are both $\varphi$ and $\psi$;
\item
	$\lnot \varphi$ is true on $w$ at $i$,
	if $\varphi$ is not.
\end{itemize}
\end{definition}

The following abbreviations are used:
$C = \tau$ stands for $(C \leqslant \tau \land C \geqslant \tau)$;
$C < \tau$ stands for $(C \leqslant \tau \land \lnot (C \geqslant \tau))$;
$C > \tau$ stands for $(C \geqslant \tau \land \lnot (C \leqslant \tau))$.

\begin{figure}[t]
	\centerline{%
	\includegraphics[scale=0.95]{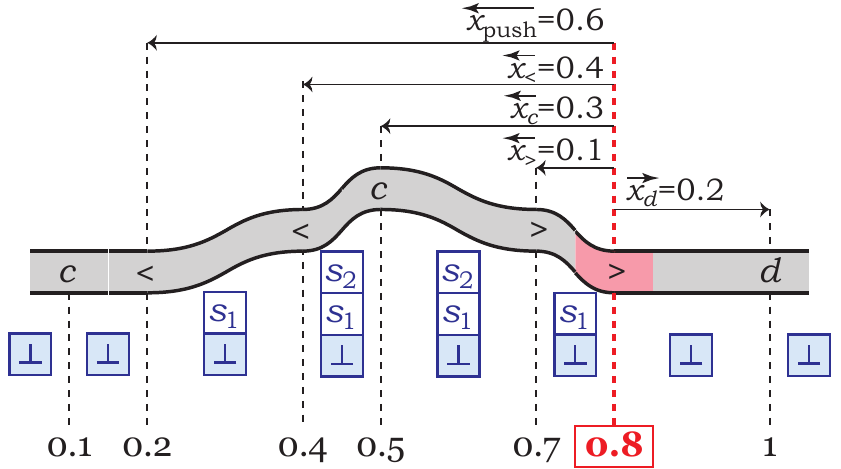}
	}
	\caption{%
		Clock values for the string
		\protect{$w=(0.1, c) (0.2, {<}) (0.4, {<}) (0.5, c) (0.7, {>}) \bm{(0.8, {>})} (1, d)$,}
		at the last right bracket,
		as in Example~\ref{timed_idpda_example}.}
	\label{f:timed_idpda_example}
\end{figure}

\begin{example}\label{timed_idpda_example}
Let $\Sigma=\Sigma_{+1} \cup \Sigma_{-1} \cup \Sigma_0$,
with $\Sigma_{+1}=\{{<}\}$, $\Sigma_{-1}=\{{>}\}$ and $\Sigma_0=\{c, d\}$, be an alphabet.
Let $w=(0.1, c) (0.2, {<}) (0.4, {<}) (0.5, c) (0.7, {>}) \bm{(0.8, {>})} (1, d)$
be a well-nested timed string over this alphabet,
illustrated in Figure~\ref{f:timed_idpda_example}.

Then, the values of the clocks at position 6 (the last right bracket)
are as follows:
$\backwardstackclock = 0.8-0.2=0.6$,
$\backwardclock{<} = 0.8-0.4=0.4$,
$\backwardclock{c} = 0.8-0.5=0.3$,
$\backwardclock{>} = 0.8-0.7=0.1$,
$\backwardclock{d}$ undefined,
$\forwardclock{<}$ undefined,
$\forwardclock{c}$ undefined,
$\forwardclock{>}$ undefined,
$\forwardclock{d} = 1-0.8=0.2$,
$\forwardstackclock$ undefined.
Accordingly, the clock constraint $\backwardstackclock > 0.1 \lor \forwardclock{c} \geqslant 0$ is true,
whereas $\backwardclock{c} > 0.1 \land \forwardclock{d} < 0.2$ is false.
\end{example}

An event-clock automaton is equipped with a finite set of such clock constraints,
and, at each step of its computation,
it knows the truth value of each of them,
and can use this information to determine its transition.
The following definition is based on Nguyen and Ogawa~\cite{NguyenOgawa}
and on Bhave et al.~\cite{BhaveDaveKrishnaPhawadeTrivedi}.
\begin{definition}\label{def:ECIDPDA}
A nondeterministic event-clock input-driven pushdown automaton (ECIDPDA)
is an octuple
$M=(\Sigma_{+1}, \Sigma_0, \Sigma_{-1}, Q, Q_0, \Gamma, \langle\delta_a\rangle_{a \in \Sigma}, F)$,
in which:
\begin{itemize}
\item
	$\Sigma = \Sigma_{+1} \cup \Sigma_{-1} \cup \Sigma_0$
	is an input alphabet split into three disjoint classes;
\item
	$Q$ is a finite set of states;
\item
	$\Gamma$ is the pushdown alphabet;
\item
	$Q_0 \subseteq Q$ is the set of initial states;
\item
	for each neutral symbol $c \in \Sigma_0$,
	the state change is described
	by a partial function $\delta_c \colon Q \times \clockconstraints{\Sigma} \to 2^Q$;
\item
	the transition function
	by each left bracket symbol ${<} \in \Sigma_{+1}$
	is $\delta_< \colon Q \times \clockconstraints{\Sigma} \to 2^{Q \times \Gamma}$,
	which, for a given current state and the truth value of clock constraints,
	provides zero or more transitions
	of the form (next state, symbol to be pushed);
\item
	for every right bracket symbol ${>} \in \Sigma_{-1}$,
	there is a partial function $\delta_> \colon Q \times (\Gamma \cup \{\bot\}) \times \clockconstraints{\Sigma} \to 2^Q$
	specifying possible next states,
	assuming that the given stack symbol is popped from the stack,
	or the stack is empty ($\bot$);
\item
	$F \subseteq Q$
	is the set of accepting states.
\end{itemize}
The domain of the transition function by each symbol must be finite.

An accepting computation of $\mathcal{A}$
on a timed string $w=(a_1, t_1) \ldots (a_n, t_n)$
is any sequence $(q_0, \alpha_0)$, $(q_1, \alpha_1)$, \ldots, $(q_n, \alpha_n)$,
with $q_0, \ldots, q_n \in Q$,
and $\alpha_0, \ldots, \alpha_n \in \Gamma^*$,
that satisfies the following conditions.
\begin{itemize}
\item
	It begins in an initial state $q_0 \in Q_0$
	with the empty stack, $\alpha_0=\epsilon$.
\item
	For each $i \in \{1, \ldots, n\}$, with $a_i=c \in \Sigma_0$,
	there exists a clock constraint $\varphi_i$ that is true on $w$ at position $i$,
	with $q_i \in \delta_c(q_{i-1}, \varphi_i)$
	and $\alpha_i=\alpha_{i-1}$.
\item
	For each $i \in \{1, \ldots, n\}$, with $a_i={<} \in \Sigma_{+1}$,
	there exists a clock constraint $\varphi_i$ that is true on $w$ at position $i$,
	with $(q_i, s) \in \delta_{<}(q_{i-1}, \varphi_i)$
	and $\alpha_i=s \alpha_{i-1}$
	for some $s \in \Gamma$.
\item
	For each $i \in \{1, \ldots, n\}$, with $a_i={>} \in \Sigma_{-1}$,
	if $\alpha_{i-1}=s \beta$ for some $s \in \Gamma$ and $\beta \in \Gamma^*$, then
	there exists a clock constraint $\varphi_i$ that is true on $w$ at position $i$,
	with $q_i \in \delta_{>}(q_{i-1}, s, \varphi_i)$
	and $\alpha_i = \beta$
\item
	For each $i \in \{1, \ldots, n\}$, with $a_i={>} \in \Sigma_{-1}$,
	if $\alpha_{i-1}=\epsilon$, then
	there exists a clock constraint $\varphi_i$ that is true on $w$ at position $i$,
	with $q_i \in \delta_{>}(q_{i-1}, \bot, \varphi_i)$
	and $\alpha_i = \epsilon$.
\item
	The computation ends in an accepting state $q_n \in F$
	with any stack contents.
\end{itemize}

The language recognized by $\mathcal{A}$,
denoted by $L(\mathcal{A})$,
is the set of all timed strings,
on which $\mathcal{A}$ has at least one accepting computation.
\end{definition}

\begin{definition} \label{def:EVIDPDA}
A nondeterministic event-clock input-driven automaton $\mathcal{A}=(\Sigma, Q, Q_0, \delta, F)$
is said to be deterministic,
if the following conditions hold.
\begin{enumerate}
\item
	There is a unique initial state: $|Q_0|=1$.
\item
	Every transition function $\delta_a$, with $a \in \Sigma_0 \cup \Sigma_{+1}$,
	satisfies $|\delta_a(q, \varphi)| \leqslant 1$
	for all $q \in Q$ and $\varphi \in \Phi(\Sigma)$,
	and whenever $\delta_a(q, \varphi)$ and $\delta_a(q, \varphi')$,
	with $\varphi \neq \varphi'$, are both non-empty,
	the clock constraints $\varphi$ and $\varphi'$
	cannot both be true at the same position of the same string.
\item
	Similarly, every transition function $\delta_>$, with ${>} \in \Sigma_{-1}$,
	satisfies $|\delta_>(q, s, \varphi)| \leqslant 1$
	for all $q \in Q$, $s \in \Gamma \cup \{\bot\}$ and $\varphi \in \Phi(\Sigma)$,
	and whenever $\delta_c(q, s, \varphi)$ and $\delta(q, s, \varphi')$,
	with $\varphi \neq \varphi'$, are both non-empty,
	the clock constraints $\varphi$ and $\varphi'$
	cannot both be true at the same position of the same string.
\end{enumerate}
\end{definition}

The first result of this paper is that nondeterministic event-clock input-driven pushdown automata
can be determinized.
Determinization results for every similar models were earlier given
by Nguyen and Ogawa~\cite{NguyenOgawa}
and by Bhave et al.~\cite{BhaveDaveKrishnaPhawadeTrivedi}.
However, their constructions relied on the method of \emph{region construction},
in which the space of clock values is discretized.
On the other hand, the construction in the present paper has the benefit of being \emph{direct},
in the sense that the transition function for a deterministic automaton
directly simulates the transitions of a nondeterministic automaton.
Later it will be proved that this easier construction is also optimal
with respect to the number of states and stack symbols.
The proposed construction is not much more difficult
than the construction for standard input-driven automata, without time.
The latter construction is used as a model,
and is recalled in the next section.

\section{Determinization of standard IDPDA without clocks}

The proposed new determinization
of nondeterministic event-clock input-driven pushdown automaton
extends the well-known determinization for standard input-driven pushdown automata,
discovered by von Braunm\"uhl and Verbeek~\cite{vonBraunmuehl_Verbeek}
and later by Alur and Madhusudan~\cite{AlurMadhusudan,AlurMadhusudan2}.

To begin with the definition,
nondeterministic input-driven automata (NIDPDA),
operating on standard strings $w=a_1 \ldots a_n$,
are defined exactly like that ECIDPDA,
with all mentions of clock constraints removed:
a transition function by a neutral symbol is $\delta_c \colon Q \to 2^Q$,
it is $\delta_{<} \colon Q \to 2^{Q \times \Gamma}$ for a left bracket
and $\delta_{>} \colon Q \times (\Gamma \cup \{\bot\}) \to 2^Q$ for a right bracket.
An NIDPDA is deterministic (DIDPDA) if each transition function gives a singleton set
for any arguments.

\begin{theorem}[von Braunm\"uhl and Verbeek~\cite{vonBraunmuehl_Verbeek}]\label{idpda_determinization_theorem}
An NIDPDA $\mathcal{A} = (\Sigma, Q, \Gamma, Q_0, \bot, [\delta_a]_{a \in \Sigma}, F)$
over an alphabet $\Sigma = \Sigma_{+1} \cup \Sigma_{-1} \cup \Sigma_0$
can be simulated by a DIDPDA
$\mathcal{B} = (\Sigma, Q', \Gamma', Q'_0, \bot, [\delta'_a]_{a \in \Sigma}, F')$,
with the set of states $Q' = 2^{Q \times Q}$,
and with the stack alphabet $\Gamma' = \Sigma_{+1} \times 2^{Q \times Q}$.
\end{theorem}
\begin{proof}
Every state $P \subseteq Q \times Q$ of $\mathcal{B}$
contains pairs of states of $\mathcal{A}$,
each corresponding to the following situation:
whenever $(p, q) \in P$,
both $p$ and $q$ are states in one of the computations of $\mathcal{A}$,
where $q$ is the state at the current position,
whereas $p$ was the state just before starting to read
the longest well-nested substring ending at the current position.

The initial state of $\mathcal{B}$,
defined as $q'_0 = \set{(q, q)}{q \in Q_0}$,
represents the behaviour of $\mathcal{A}$ on the empty string,
which begins its computation in an initial state,
and remains in the same state.
The set of accepting states
reflects all computations of $\mathcal{A}$ ending in an accepting state.
\begin{equation*}
	F'
		=
	\set{P \subseteq Q \times Q}{
	\text{there is a pair } (p, q) \in P \text{ with } q \in F}
\end{equation*}
The transition functions $\delta'_a$, with $a \in \Sigma$,
are defined as follows.
\begin{itemize}
\item
	For a neutral symbol $c \in \Sigma_0$ and a state $P \in Q_B$,
	the transition
	$\delta'_c(P) = \{ (p, q') \mid \exists (p, q) \in P : q' \in \delta_c(q) \}$
	directly simulates one step of $A$
	in all currently traced computations.
\item
	On a left bracket ${<} \in \Sigma_{+1}$,
	the transition in a state $P \in Q_B$ is
	$\delta'_{<}(P) = (P', ({<}, P))$, where
	\begin{equation*}
		P'
			=
		\{ (q', q') \mid (\exists (p, q) \in P) 
			(\exists \gamma \in \Gamma) : 
			(q', \gamma) \in \delta_{<}(q) \}.
	\end{equation*}
	Thus, $B$ pushes the current context of the simulation onto the stack,
	along with the current left bracket,
	and starts the simulation afresh at the next level of brackets,
	where it will trace the computations
	from all states $q'$ reachable by $A$ at this point.
\item
	For a right bracket ${>} \in \Sigma_{-1}$
	and a state $P' \subseteq Q_B$,
	the automaton pops a stack symbol $({<}, P) \in \Gamma_B$
	containing a matching left bracket
	and the context of the previous simulation.
	Then, each computation in $P$
	is continued by simulating the transition by the left bracket,
	the behaviour inside the brackets stored in $P'$,
	and the transition by the right bracket.
	\begin{equation*}
	\delta'_{>}(P', ({<}, P))
		=
	\{ (p, q'')
	\mid
	(\exists (p, q) \in P)
	(\exists (p', q') \in P')
	(\exists s \in \Gamma):
	(p', s) \in \delta_{<}(q),
	q'' \in \delta_{>}(q', s)
	\}.
	\end{equation*}
\item
	For an unmatched right bracket ${>} \in \Sigma_{-1}$,
	the transition in a state $P \in Q_B$
	advances all currently simulated computations of $A$
	in the same way as for a neutral symbol:
	$\tau_{>}(P, \bot)
		=
	\{ (p, q')
	\mid
	\exists (p, q) \in P:
	q' \in \delta_{>}(q', \bot)
	\}$.
\end{itemize}
The correctness of the construction
can be proved by induction on the bracket structure
of an input string.
\end{proof}

This construction is asymptotically optimal:
as proved by Alur and Madhusudan~\cite{AlurMadhusudan,AlurMadhusudan2},
$2^{\Theta(n^2)}$ states are necessary in the worst case.
Okhotin, Piao and Salomaa~\cite[Thm.~3.2]{OkhotinPiaoSalomaa} refined this estimation
to show that in the worst case
a deterministic automaton also requires $2^{\Theta(n^2)}$ stack symbols.

\section{Direct determinization of event-clock IDPDA}\label{section_ECIDPDA_determinization}

The determinization construction in Theorem~\ref{idpda_determinization_theorem}
shall now be extended to handle event-clock input-driven automata.

The original untimed construction in Theorem~\ref{idpda_determinization_theorem}
is based upon considering the original automaton's behaviour
on a left bracket and on a matching right bracket at the same time,
while reading the right bracket.
In this way, the stack symbol pushed while reading the left bracket
is matched to the symbol popped while reading the right bracket,
and all possible computations of this kind can be considered at once.
However, in the event-clock case,
the nondeterministic decisions made on a left bracket
are based upon the clock values at that time,
and if the simulation of thise decisions were deferred
until reading the matching right bracket,
then those clock values would no longer be available.
Since event-clock automata cannot manipulate clock values explicitly,
they, in particular, cannot push the clock values onto the stack
for later use.
However, what can be done
is to \emph{test all elementary clock constraints while reading the left bracket},
store their truth values in the stack,
and later, upon reading the right bracket,
use this information to simulate the behaviour of the original automaton
on the left bracket.
This idea is implemented in the following construction,
which uses the same set of states as in Theorem~\ref{idpda_determinization_theorem},
but more complicated stack symbols.

\begin{theorem}\label{determinization_without_stack_prediction_constraints_theorem}
Let $\mathcal{A}=(\Sigma_{+1}, \Sigma_0, \Sigma_{-1}, Q, Q_0, \Gamma, \langle\delta_a\rangle_{a \in \Sigma}, F)$
be a nondeterministic event-clock input-driven automaton.
Let $\Psi$ be the set of atomic constraints used in its transitions.
Then there exists a deterministic event-clock input-driven automaton
with the set of states $Q'=2^{Q \times Q}$,
and with the pushdown alphabet $\Gamma'=2^{Q \times Q} \times \Sigma_{+1} \times 2^{\Psi}$,
which recognizes the same set of timed strings as $\mathcal{A}$.
\end{theorem}
\begin{proof}
States of the deterministic automaton $\mathcal{B}$
are again pairs $(p, q) \in P$,
which means, as in Theorem~\ref{idpda_determinization_theorem},
that there is a computation of the original automaton $\mathcal{A}$
on the longest well-nested suffix of the input,
which begins in the state $p$ and ends in the state $q$.

The initial state of $\mathcal{B}$ is $q'_0 = \set{(q_0, q_0)}{q \in Q_0}$.

For a \textbf{neutral symbol} $c \in \Sigma_0$ and a state $P \in Q'$,
the transition $\delta'_c(P)$
advances all current computations traced in $P$
by the next symbol $c$.
Each computation continues by its own transition,
which requires a certain clock constraint to be true.
Whether each clock constraint $\varphi \in \clockconstraints{\Sigma}$ is true or false,
can be deduced from the truth assignment to the atomic constraints.
In other words, for every set of atomic constraints $S \subseteq \Psi$ assumed to be true,
$\varphi$ is either true or false under the assignment $S$.
Also let $\xi_S = \bigwedge_{C \in S} C \land \bigwedge_{C \in \Psi \setminus S} \lnot C$
be a clock constraint asserting that among all atomic constraints,
exactly those belonging to $S$ are true.
Then, for every set $S$, the new automaton has the following transition.
\begin{equation*}
	\delta_c(P, \xi_S) = \set{(p, q')}{\exists (p, q) \in P, \: \exists \varphi:
		q' \in \delta_c(q, \varphi), \; \varphi \text{ is true under } S}
\end{equation*}

On a \textbf{left bracket} ${<} \in \Sigma_{+1}$,
the transition of $\mathcal{B}$ in a state $P \in Q'$
pushes the current context of the simulation onto the stack,
and starts the simulation afresh at the next level of brackets,
where it will trace the computations
beginning in different states $r \in Q$.
A computation in a state $r$
is started only if any computations of $\mathcal{A}$ actually reach that state.
In addition, $\mathcal{B}$ pushes the current left bracket ($<$),
as well as the truth value of all atomic constraints at the present moment, $S \subseteq \Psi$.
This is done in the following transitions,
defined for every set of atomic constraints $S \subseteq \Psi$.
\begin{multline*}
	\delta'_{<}(P, \xi_S) = \big(
		\set{(r,r)}{\exists (p, q) \in P, \: \exists \varphi \in \clockconstraints{\Sigma}):
			\varphi \text{ is true under } S, \: r \in \delta_<(q)
		}, \;
		(P, {<}, S)
	\big)
\end{multline*}
All these data are only \emph{stored} in the stack;
at present,
the transitions of $\mathcal{A}$ on this left bracket ($<$)
are considered only to the extent of determining all reachable states $r$.
If a matching right bracket ($>$) is eventually read,
then the computations of $\mathcal{A}$ reflected in $P$
shall be simulated further at that moment:
then $\mathcal{B}$ shall pop $(P, {<}, S)$ from the stack
and reconstruct what has happened to each of the computations of $\mathcal{A}$
at this point and further on.
On the other hand, if this left bracket ($<$) is unmatched,
then the acceptance shall be determined
on the basis of the computations traced on the nested level of brackets.

When $\mathcal{B}$ encounters a \textbf{matched right bracket} ${>} \in \Sigma_{-1}$
in a state $P' \subseteq Q \times Q$,
it pops a stack symbol $(P, {<}, S) \in \Gamma'$
containing the matching left bracket (${<} \in \Sigma_{+1}$),
the data on all computations on the current level of brackets
simulated up to that bracket ($P \subseteq Q \times Q$),
and the truth value of all atomic clock constraints
at the moment of reading that bracket ($S \subseteq \Psi$).

\begin{figure}[t]
	\centerline{%
	\includegraphics[scale=0.95]{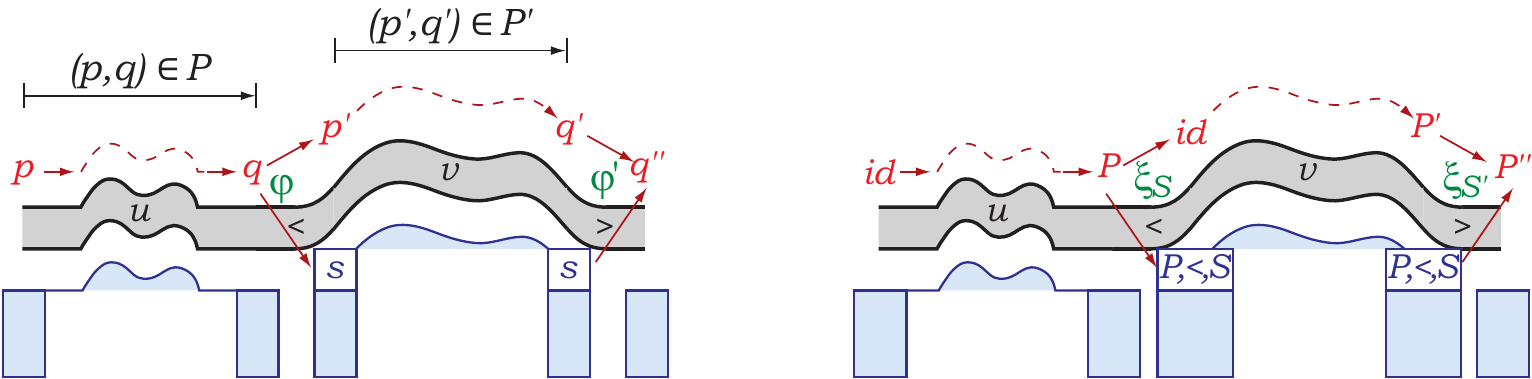}
	}
	\caption{(left) A computation of a nondeterministic event-clock IDPDA;
		(right) Its simulation by a deterministic event-clock IDPDA.}
	\label{f:timed_idpda_determinization}
\end{figure}

Then, each computation in $P$
is continued by simulating the transition by the left bracket ($<$),
the behaviour inside the brackets stored in $P'$,
and the transition by the right bracket ($>$).
Let $u{<}v{>}$ be the longest well-nested suffix of the string read so far.
Every computation of $\mathcal{A}$ on $u$, which begins in a state $p$ and ends in a state $q$,
is represented by a pair $(p, q)$.
Upon reading the left bracket ($<$),
the automaton $\mathcal{A}$
makes a transition to a state $p'$, pushing a stack symbol $s$,
along with checking a clock constraint $\varphi$.
The automaton $\mathcal{B}$ can now check the same clock constraint
by using the set of $S$ of atomic clock constraints
that held true at the earlier left bracket ($<$).
For every set of atomic constraints $S' \subseteq \Psi'$,
the following transition is defined.
\begin{align*}
	\delta'_>(P', (P, {<}, S), \xi_{S'})
		=
	\big\{\: (p, q'') \: \big| \:
	(\exists (p, q) \in P)
	(\exists (p', q') \in P')
	(\exists s \in \Gamma)
	(\exists \varphi, \varphi' \in \clockconstraints{\Sigma}) &: \\
	\varphi \text{ is true under } S, \:
	(p', s) \in \delta_{<}(q, \varphi), \;
	\varphi' \text{ is true under } S', \:
	q'' \in \delta_{>}(q', s, \varphi')
	&\big\}
\end{align*}

When $\mathcal{B}$ reads an \textbf{unmatched right bracket} ${>} \in \Sigma_{-1}$	
while in a state $P \subseteq Q \times Q$,
it continues the existing computations
on the new bottom level of brackets.
\begin{equation*}
	\delta_>(P, \bot, \xi_S) = \set{(p', p')}{\exists (p, q) \in P, \: \exists \varphi:
		p' \in \delta_>(q, \bot, \varphi), \; \varphi \text{ is true under } S}
\end{equation*}

The set of \textbf{accepting states}
reflects all computations of $\mathcal{A}$ ending in an accepting state.
\begin{equation*}
	F'
		=
	\set{P \subseteq Q \times Q}{
		\text{there is a pair } (p, q) \in P, \text{ with } q \in F}
\end{equation*}

A formal correctness claim for this construction reads as follows.

\begin{claim}
Let $uvw$ be a timed string,
where $v$ is the longest well-nested suffix of $uv$,
and let $P \subseteq Q \times Q$ be the state reached by $\mathcal{B}$ on $uvw$
after reading $uv$.
Then a pair $(p, p')$ is in $P$
if and only if
there is a computation of $\mathcal{A}$ on $uvw$
that passes through the state $p$ right after reading $u$,
and later, after reading the following $v$,
enters the state $p'$.
\end{claim}

The claim can be proved by induction on the bracket structure
of an input string.
\end{proof}

It is interesting to note that the above determinization construction
does not rely on the exact form of clock constraints:
the resulting deterministic automaton uses any kind of constraints
used by the original nondeterministic automaton,
and only communicates the results through the stack in the form of Boolean values.
The same construction would apply verbatim
for any kind of contraints expressed in the model:
these could be any constraints
mapping any pair of a timed string $(a_1, t_1) \ldots (a_\ell, t_\ell)$
and a position $i \in \{1, \ldots, \ell\}$
to true or false.

\section{A lower bound on the determinization complexity}\label{section_ECIDPDA_determinization_lower_bound}

The timed determinization construction
in Theorem~\ref{determinization_without_stack_prediction_constraints_theorem}
produces $2^{n^2}$ states and $2^{n^2+k}$ stack symbols,
where $n$ is the number of states in the nondeterministic automaton
and $k$ is the number of atomic clock constraints.
It shall now be proved that this construction is asymptotically optimal.
The following theorem, proved in the rest of this section,
is a timed extension of a result by Okhotin, Piao and Salomaa~\cite[Thm.~3.2]{OkhotinPiaoSalomaa}.

\begin{theorem}\label{event_clock_idpda_determinization_lower_bound_theorem}
For every $n$ and for every $k$,
there is an $O(n)$-state nondeterministic ECIDPDA
over an alphabet of size $k+O(1)$,
with $nk$ stack symbols and $k$ atomic constraints referring only to symbol history clocks,
such that every deterministic ECIDPDA recognizing the same timed language
must have at least $2^{n^2}$ states and at least $2^{n^2-O(n)+k}$ stack symbols.
\end{theorem}

The automaton is defined over the following alphabet:
$\Sigma_{+1}=\{{<}\}$,
$\Sigma_{-1}=\{{>}\}$,
$\Sigma_0=\{a, b, c, \#\} \cup \set{e_i}{1 \leqslant i \leqslant k}$.

The problem solved by the automaton requires some notation to express.
For a set of pairs $R=\{(i_1, j_1), \ldots, (i_\ell, j_\ell)\} \subseteq \{0, \ldots, n-1\}^2$,
let $u_R \in \{a,b,\#\}$ be the string that lists all pairs in $R$ in the lexicographical order,
under the following encoding.
\begin{equation*}
	u_R = \#a^{i_1}b^{j_1} \, \#a^{i_2}b^{j_2} \ldots \#a^{i_\ell}b^{j_\ell}
\end{equation*}
For every set of symbols $X = \{e_{i_1}, \ldots, e_{i_\ell}\} \subseteq \{e_1, \ldots, e_k\}$,
let $v_X=e_1 \ldots e_k e_{i_1} \ldots e_{i_\ell}$
be the string that first lists all the symbols in $\{e_1, \ldots, e_\ell\}$,
and then only the symbols in $X$.

Now, let $m \geqslant 1$ be the number of levels in the string to be constructed,
let $s_1, \ldots, s_m, s_{m+1} \in \{0, \ldots, n-1\}^2$ be numbers,
let $R_1, \ldots, R_m \subseteq \{0, \ldots, n-1\}^2$ be relations,
and let $X_1, Y_1, \ldots, X_m, Y_m \subseteq \{e_1, \ldots, e_\ell\}$ be $2m$ sets of symbols.
This information is encoded in the following string.
\begin{equation*}
	w
		=
	\underbrace{v_{X_1} {<} u_{R_1} v_{X_2} {<} u_{R_2} \ldots
	v_{X_m} {<} u_{R_m}}_{w_1}
	\underbrace{c^{s_{m+1}} v_{Y_m} {>} c^{s_m}
	\ldots v_{Y_2} {>} c^{s_2} v_{Y_1} {>} c^{s_1}}_{w_2}
\end{equation*}
The string is extended to a timed string by supplying time values
with the following property:
in each string $v_{X_i}$, its first $k$ symbols
occur more than 1 time unit earlier than the subsequent left bracket ($<$),
whereas its remaining symbols representing the elements of $X_i$
occur less than 1 time unit earlier than the left bracket;
similarly, in each string $v_{Y_i}$, its first $k$ symbols
occur more than 1 time unit earlier than the next right bracket ($>$),
while its remaining symbols occur less than 1 time unit earlier than the bracket.
This allows an event-clock automaton
to \emph{see} the set $X_i$ using clock constraints while reading the left bracket ($<$),
and to see $Y_i$ while at the right bracket ($>$).\footnote{%
	Some further technical extensions to the encoding are necessary
	to make sure that the automaton cannot see anything else using any clock constraints.
	It is sufficient to list all symbols at predefined moments of time before and after every substring,
	so that no checks based on clock constraints
	could reveal anything on the contents of these substrings.
	To keep the notation simple,
	these details are omitted in the present version of this paper.
}

A timed string is said to be \emph{well-formed} if it is defined as above,
for some $m$, $s_i$, $R_i$, $X_i$ and $Y_i$.
A well-formed string is said to be \emph{valid}, if the following conditions hold.
\begin{itemize}
\item
	First, $(s_i, s_{i+1}) \in R_i$ for each $i$,
	that is, every two subsequent numbers given in the suffix $w_2$
	must be listed as $\#a^s b^t$ in the encoding $u_{R_i}$
	at the corresponding level of brackets.
\item
	Secondly, $X_i \cap Y_i \neq \emptyset$ for each $i$,
	that is, there exists a symbol $e \in \{e_1, \ldots, e_k\}$
	that occurs less than 1 time unit before the left bracket ($<$),
	and later occurs again less than 1 time unit before the right bracket ($>$).
\end{itemize}

\begin{figure}[t]
	\centerline{%
	\includegraphics[scale=0.95]{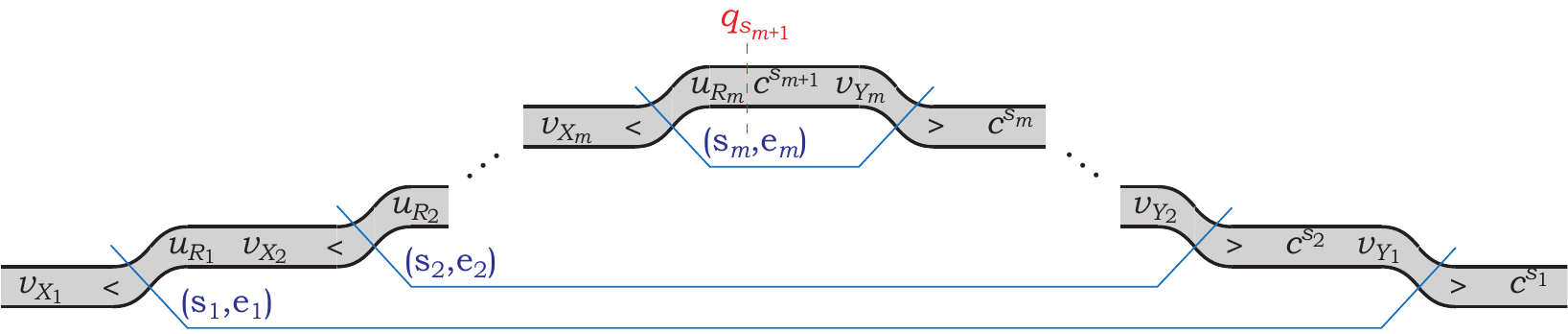}
	}
	\caption{A nondeterministic event-clock IDPDA checking the validity of a well-formed string.}
	\label{f:timed_idpda_lower_bound}
\end{figure}

\begin{lemma}\label{event_clock_idpda_determinization_lower_bound_theorem__nondeterministic_automaton_lemma}
For every $n$ and $k$,
there exists a nondeterministic ECIDPDA
using $O(n)$ states, $nk$ stack symbols and $k$ clock constraints,
which accepts every valid well-formed string
and does not accept any invalid well-formed string.
\end{lemma}
\begin{proof}
This automaton operates as follows.
First, it skips the symbols of $v_{X_1}$.
At the first transition upon the left bracket ($<$),
it nondeterministically guesses the number $s_1 \in \{0, \ldots, n-1\}$
and a symbol $e_1$ in $X_1$ (checked by a clock constraint $\backwardclock{e_1} < 1$);
then it pushes the pair $(s_1, e_1)$ onto the stack
and enters a state in which it remembers the number $s_1$
Inside the brackets, the automaton skips some prefix of $u_{R_1}$
until eventually, at some separator $\#$,
it nondeterministically decides to find $s_1$ here.
If this separator is followed by a substring other than $a^{s_1}$,
the automaton rejects;
otherwise, it forgets the number $s_1$
and reads a number $s_2$ from the following substring $b^{s_2}$.
Then it skips the rest of $u_{R_1}$ and the whole string $v_{X_2}$ while remembering the number $s_2$.
Upon seeing the next left bracket ($<$),
the automaton nondeterministically guesses a symbol $e_2$ in $X_2$ (and verifies it by a clock constraints)
and pushes the pair $(s_2, e_2)$ onto the stack,
entering the next level of brackets in a state in which it remembers $s_2$.

The process continues, until the automaton
eventually finishes reading the first half of the input ($w_1$).
At this time, it has pairs $(s_1, e_1)$, \ldots, $(s_m, e_m)$ in the stack
and a number $s_{m+1}$ in the current state,
which satisfy two conditions:
first, $(s_i, s_{i+1}) \in R_i$ for all $i$,
and secondly, $e_i \in X_i$ for each $i$.
In the rest of the computation, while reading $w_2$,
the automaton shall verify that the strings $c^{s_{m+1}}$, $c^{s_m}$, \ldots, $c^{s_1}$
encode exactly these numbers,
and that each sets $Y_i$ encoded in a string $v_{Y_i}$
contains the corresponding symbol $e_i$.

As the automaton starts reading the string $c^{s_{m+1}}$ in the state $s_{m+1}$,
it checks this single number.
Then it skips the substring $v_{Y_m}$.
Upon reading the right bracket ($>$),
the automaton pops the pair $(s_m, e_m)$
and uses clock constraints to verify that $e_m$ is in $Y_m$.
The number $s_m$ is read into the current state,
and the automaton proceeds to check the rest of the numbers and symbols in the same way.

If any checks fail, the automaton rejects immediately.
If all checks are passed, the automaton finishes reading the string and accepts.
\end{proof}

\begin{lemma}
For every $n$ and $k$,
every deterministic ECIDPDA
that accepts every valid well-formed string
and does not accept any invalid well-formed string
must have at least $2^{n^2}$ states.
\end{lemma}
\begin{proof}
This is a standard argument, which does not use clocks at all.
It is sufficient to use one-level well-formed strings,
defined for two numbers $s, t \in \{0, \ldots, n-1\}^2$,
one relation $R \subseteq \{0, \ldots, n-1\}^2$
and two sets of symbols $X, Y, \subseteq \{e_1, \ldots, e_\ell\}$.
\begin{equation*}
	w
		=
	\underbrace{v_{X} {<} u_{R}}_{w_1} \underbrace{c^{t} v_{Y} {>} c^{s}}_{w_2}
\end{equation*}
At the moment of reading the left bracket ($<$),
a deterministic automaton knows only the set $X$ and nothing else.
The symbol pushed at this moment does not depend on the relation $R$.
Then, after finishing reading $w_1$
the automaton has to remember the entire set $R$ in its internal state,
so that it could later check that the pair $(s, t)$ is in $R$.

Suppose that the automaton has fewer than $2^{n^2}$ states.
Then, there exist two distinct relations, $R$ and $R'$,
with $(s, t) \in R \setminus R'$,
for which the automaton,
in its computations on the valid string 
$w=v_{\{e_1\}} {<} u_R c^t v_{\{e_1\}} {>} c^s$
and on the invalid string
$w'=v_{\{e_1\}} {<} u_{R'} c^t v_{\{e_1\}} {>} c^s$,
enters the same state before reading $c_t$.
Then the automaton either accepts both $w$ and $w'$
or rejects both strings,
which is a contradiction.
\end{proof}

\begin{lemma}\label{event_clock_idpda_determinization_lower_bound_theorem__stack_symbols_lemma}
For every $n$ and $k$,
every deterministic ECIDPDA
that accepts every valid well-formed string
and does not accept any invalid well-formed string
must have at least $2^{n^2-O(n)+k}$ stack symbols.
\end{lemma}
\begin{proof}
The proof is modelled on the proof
by Okhotin, Piao and Salomaa~\cite[Lemma~3.4]{OkhotinPiaoSalomaa},
with the clock constraints added.

The argument uses binary relations that are both left-total and right-total:
that is, relations $R \subseteq \{0, \ldots, n-1\}^2$ in which,
for every $x \in \{0, \ldots, n-1\}$,
there is an element $y$ with $(x, y) \in R$,
and, symmetrically, for every $y$,
there is an element $x$ with $(x, y) \in R$.
There are at least $2^{n^2}-2n \cdot 2^{n(n-1)}$ such relations,
hence their number is estimated as $2^{n^2-O(n)}$.

Fix the number of levels $m \geqslant 1$,
and let $R_1, \ldots, R_m \subseteq \{0, \ldots, n-1\}^2$ be left- and right-total relations,
and let $X_1, \ldots, X_m \subseteq \{e_1, \ldots, e_\ell\}$ be non-empty sets of symbols.
These parameters define the first part $w_1$ of a well-formed string.
It is claimed that, after reading $w_1$,
a deterministic automaton somehow has to store
all relations $R_1, \ldots, R_m$ and all sets $X_1, \ldots, X_m$
in the available memory: that is, in $m$ stack symbols and in one internal state.

Suppose that, for some 
$R_1, \ldots, R_m, R'_1, \ldots, R'_m \subseteq \{0, \ldots, n-1\}^2$
and $X_1, \ldots, X_m, X'_1, \ldots, X'_m \subseteq \{e_1, \ldots, e_\ell\}$,
with $(R_1, \ldots, R_m, X_1, \ldots, X_m) \neq (R'_1, \ldots, R'_m, X'_1, \ldots, X'_m)$,
the automaton, after reading the corresponding first parts $w_1$ and $w'_1$,
comes to the same state with the same stack contents.
\begin{align*}
	w_1
		&=
	v_{X_1} {<} u_{R_1} v_{X_2} {<} u_{R_2} \ldots
	v_{X_m} {<} u_{R_m}
\\
	w'_1
		&=
	v_{X'_1} {<} u_{R'_1} v_{X'_2} {<} u_{R'_2} \ldots
	v_{X'_m} {<} u_{R'_m}
\end{align*}

First, as in the argument by Okhotin, Piao and Salomaa~\cite[Lemma~3.4]{OkhotinPiaoSalomaa},
assume that these parameters differ in an $i$-th relation, with $(s, t) \in R_i \setminus R'_i$.
Let $s_i=s$. Since all relations $R_{i-1}, \ldots, R_1$ are right-total,
there exists a sequence of numbers $s_{i-1}, \ldots, s_1$,
with $(s_j, s_{j+1}) \in R_j$ for all $j \in \{1, \ldots, i-1\}$.
Similarly, let $s_{i+1}=t$. Since the relations $R_{i+1}, \ldots, R_m$ are left-total,
there is a sequence $s_{i+2}, \ldots, s_{m+1}$,
with $(s_j, s_{j+1}) \in R_j$ for all $j \in \{i+1, \ldots, m\}$.
Construct the following continuation for $w_1$ and $w'_1$.
\begin{equation*}
	w_2
		=
	c^{s_{m+1}} v_{X_m} {>} c^{s_m}
	\ldots v_{X_2} {>} c^{s_2} v_{X_1} {>} c^{s_1}
\end{equation*}
The concatenation $w_1 w_2$ is then well-formed and valid,
whereas the concatenation $w'_1 w_2$ is well-formed and invalid,
because $(s_i, s_{i+1}) \notin R'_i$.
Since the automaton either accepts both or rejects both,
a contradiction is obtained.

Now assume that the prefixes $w_1$ and $w'_1$
use the same relations $R_1, \ldots, R_m$
and differ in an $i$-th set, with $e \in X_i \setminus X'_i$.
Since all relations are left-total,
there exists a sequence of numbers $s_1, \ldots, s_m, s_{m+1}$,
with $(s_j, s_{j+1}) \in R_j = R'_j$ for all $j \in \{1, \ldots, m\}$.
This time, the continuation includes the sequence of numbers
and takes all sets $X_j$ from $w_1$,
except for $X_i$, which is replaced by $\{e\}$.
\begin{equation*}
	w_2
		=
	c^{s_{m+1}} v_{X_m} {>} c^{s_m}
	\ldots
	v_{X_{i+1}} {>} c^{s_{i+1}}
	v_{\{e\}} {>} c^{s_i}
	v_{X_{i-1}} {>} c^{s_{i-1}}
	\ldots
	v_{X_1} {>} c^{s_1}
\end{equation*}
Then, both concatenations $w_1 w_2$ and $w'_1 w_2$ are well-formed.
However, the concatenation $w_1 w_2$ is valid,
whereas $w'_1 w_2$ is invalid, because $X'_i \cap \{e\}=\emptyset$.
But the automaton again either accepts both concatenations or rejects both of them,
which is a contradiction.

This shows that, for each $m \geqslant 1$,
the automaton must be able to reach
at least $(2^{n^2}-2n \cdot 2^{n(n-1)})^m (2^k-1)^m$ distinct configurations
after reading different strings of the given form.
Let $Q$ be the automaton's set of states
and let $\Gamma$ be its stack alphabet.
Then the following inequality must hold for every $m$.
\begin{equation*}
	|\Gamma|^m \cdot |Q| \geqslant (2^{n^2}-2n \cdot 2^{n(n-1)})^m (2^k-1)^m
\end{equation*}
Taking the $m$-th root of both sides yields the next inequality.
\begin{equation*}
	|\Gamma| \cdot \sqrt[m]{|Q|} \geqslant (2^{n^2}-2n \cdot 2^{n(n-1)})(2^k-1)
\end{equation*}
Since $\sqrt[m]{|Q|} < 2$ for $m$ large enough,
this proves the desired lower bound on the number of stack symbols.
\begin{equation*}
	|\Gamma| \geqslant 2^{n^2-O(n)+k}
\end{equation*}
\end{proof}

The proof of Theorem~\ref{event_clock_idpda_determinization_lower_bound_theorem}
follows from Lemmata~\ref{event_clock_idpda_determinization_lower_bound_theorem__nondeterministic_automaton_lemma}--\ref{event_clock_idpda_determinization_lower_bound_theorem__stack_symbols_lemma}.

\section{Improved determinization}

Another determinization construction given below
additionally eliminates all references
to the stack prediction clock ($\forwardstackclock$).
If the input string is well-nested,
such constraints could be handled within the construction
in Theorem~\ref{determinization_without_stack_prediction_constraints_theorem}:
whenever the nondeterministic automaton reads a left bracket ($<$)
while checking such constraints,
the simulating deterministic automaton
shall defer the verification of these constraints
until the matching right bracket ($>$).
However, if a left bracket ($<$) turns out to be unmatched,
then the constraint verification cannot be thus deferred,
and the construction has to be augmented
with extra states to handle this possibility.

\begin{theorem}
Let $\mathcal{A}=(\Sigma_{+1}, \Sigma_0, \Sigma_{-1}, Q, Q_0, \Gamma, \langle\delta_a\rangle_{a \in \Sigma}, F)$
be a nondeterministic event-clock input-driven automaton,
let $\Psi_0$ be the set of all atomic set prediction constraints used in its transitions,
and let $\Psi$ be the set of all other atomic constraints used in its transitions.
Then there exists a deterministic event-clock input-driven automaton
with the set of states $Q'=2^{Q \times Q} \times 2^Q$
and with the pushdown alphabet $\Gamma'=2^{Q \times Q} \times 2^Q \times \Sigma_{+1} \times 2^{\Psi}$,
which never uses the stack prediction clock ($\forwardstackclock$),
and recognizes the same language.
\end{theorem}
\begin{proof}
This time, the states of $\mathcal{B}$
are pairs $(P, R)$,
with $P \subseteq Q \times Q$ and $R \subseteq Q$.
The set $P$ is constructed in generally the same way
as in Theorem~\ref{determinization_without_stack_prediction_constraints_theorem},
with a few changes needed to eliminate all references
to the stack prediction clock ($\forwardstackclock$).
The set $R$ contains all states
reached by any computations of $\mathcal{A}$ at this point,
under the assumption that
\emph{none of the stack symbols currently in the stack shall ever be popped},
that is, all the corresponding left brackets
are unmatched.
If the end of the string is reached,
this confirms the assumption,
and acceptance can be determined based on $R$.
On the other hand,
if the top stack symbol is ever popped,
then all the data collected in $R$ are invalid
and shall be discarded.

The initial state is $q'_0 = \big(\set{(q_0, q_0)}{q \in Q_0}, \{q_0\}\big)$.

On a \textbf{neutral symbol} $c \in \Sigma_0$,
the transition $\delta'_c(P, R)$,
for $P \subseteq Q \times Q$ and $R \subseteq Q$,
advances all the computations in $P$ and in $R$ by $c$.
No stack prediction constraints are involved.
For every set of atomic constraints $S \subseteq \Psi$ assumed to be true,
the new automaton has the following transition.
\begin{align*}
	\delta_c(P, R, \xi_S) = \big(&\set{(p, q')}{\exists (p, q) \in P, \: \exists \varphi:
		q' \in \delta_c(q, \varphi), \; \varphi \text{ is true under } S}, \\
		&\set{r'}{\exists r \in R, \: \exists \varphi:
		r' \in \delta_c(r, \varphi), \; \varphi \text{ is true under } S}\big)
\end{align*}

On a \textbf{left bracket} ${<} \in \Sigma_{+1}$,
the transition $\delta'_{<}(P, R)$,
for $P \subseteq Q \times Q$ and $R \subseteq Q$,
pushes the current context of the simulation onto the stack.
At the next level of brackets,
it starts a new simulation in the first component of the state,
whereas in the second component,
the computations in $R$ are continued
under the assumption that the left bracket ($<$) being read is unmatched.
The following transition
is defined for every set of atomic constraints $S \subseteq \Psi$ assumed to be true,
with all atomic set prediction constraints assumed to be false.
\begin{align*}
	\delta'_{<}((P, R), \xi_S) = \Big[&\big(
		\set{(p',p')}{p' \in Q}, \\
		&\:\:\set{r'}{\exists r \in R, \: \exists \varphi \in \clockconstraints{\Sigma}):
			\varphi \text{ is true under } S, \: r' \in \delta_<(r)
		}\big), \\
		&(P, R, {<}, S)
	\Big]
\end{align*}

On a \textbf{matched right bracket} ${>} \in \Sigma_{-1}$,
assume that $\mathcal{B}$ is in a state $(P', R')$
and pops a stack symbol $(P, R, {<}, S) \in \Gamma'$.
For each transition, let $S' \subseteq \Psi$
be the set of all atomic clock constraints
assumed to be true at the present moment.
Under this assumption, the set $\widetilde{S} \subseteq \Psi_0$
of stack prediction constraints
that were valid at the matching left bracket ($<$)
can be determined from the symmetric stack history constraints in $S'$
by setting
$\widetilde{S}=\set{\forwardstackclock \mathop{\mathrm{op}} \tau}{(\backwardstackclock \mathop{\mathrm{op}} \tau) \in S', \: \mathrm{op} \in \{{\leqslant}, {\geqslant}\}}$.
In all other respects,
the set of pairs in the new state
is determined by the same rules
as in Theorem~\ref{determinization_without_stack_prediction_constraints_theorem}.

Turning to the second component in the new state,
the set $R'$ is discarded,
because it is valid only under the assumption
that no stack symbols shall be popped,
whereas the present transition is an evidence to the contrary.
Therefore, $\mathcal{B}$ takes the earlier set $R$
and continues all the computations traced therein.
\begin{align*}
	\delta'_>((P', R'), (P, R, {<}, S), \xi_{S'})
		=
	\Big[\big\{\: (p, q'') \: \big| \hspace*{5cm} \\
	(\exists (p, q) \in P)
	(\exists (p', q') \in P')
	(\exists s \in \Gamma)
	(\exists \varphi, \varphi' \in \clockconstraints{\Sigma}) &: \\
	\varphi \text{ is true under } S \cup \widetilde{S}, \:
	(p', s) \in \delta_{<}(q, \varphi)&, \\
	\varphi' \text{ is true under } S', \:
	q'' \in \delta_{>}(q', s, \varphi')
	&\big\}, \\
	\big\{\: r''' \: \big| \:
	(\exists r \in R)
	(\exists (r', r'') \in P')
	(\exists s \in \Gamma)
	(\exists \varphi, \varphi' \in \clockconstraints{\Sigma}) &: \\
	\varphi \text{ is true under } S, \:
	(r', s) \in \delta_{<}(r, \varphi)&, \\
	\varphi' \text{ is true under } S', \:
	r''' \in \delta_{>}(r'', s, \varphi')
	&\big\}
	\Big]
\end{align*}

On an \textbf{unmatched right bracket} ${>} \in \Sigma_{-1}$	
if $\mathcal{B}$ is in a state $(P, R)$,
then it discards $P$ and starts new computations
on the new bottom level of brackets,
whereas the computations represented by $R$
are continued into that level.
\begin{align*}
	\delta_>((P, R), \bot, \xi_S) = \big(&\set{(p', p')}{p' \in Q}, \\
		&\set{r'}{\exists r \in R, \: \exists \varphi:
		r' \in \delta_>(r, \bot, \varphi), \; \varphi \text{ is true under } S}
	\big)
\end{align*}

A state $(P, R)$ is set to be \textbf{accepting}
if $R$ contains at least one accepting state of $\mathcal{A}$.
\begin{equation*}
	F'
		=
	\set{(P, R)}{R \cap F \neq \emptyset}
\end{equation*}
Indeed, if $\mathcal{B}$ finishes reading the input string
in a state $(P, R)$,
then all stack symbols currently in the stack
shall never be popped,
and therefore $R$ is the set of all states,
in which $\mathcal{A}$ may finish reading this string.

\begin{claim}
On a timed string $uvw$,
where $v$ is the longest well-nested suffix of $uv$,
after reading $uv$,
the automaton $\mathcal{B}$ reaches a state $(P, R)$,
with the following values of $P \subseteq Q \times Q$ and $R \subseteq Q$.
The set $P$ contains a pair $(p, p')$
if and only if
there is a computation of $\mathcal{A}$ on $uvw$
that passes through the state $p$ right after reading $u$,
and later, after reading the following $v$,
enters the state $p'$.
Under the assumption that all left brackets unmatched in $u$
are unmatched in $uvw$,
the set $R$ contains a state $r$
if and only if
there is a computation of $\mathcal{A}$ on $uvw$
that reaches the state $r$ after reading $uv$.
If the assumption on the unmatched brackets does not hold,
then the value of $R$ is undefined.
\end{claim}

This claim is proved by the same kind of induction
as in the proof of Theorem~\ref{determinization_without_stack_prediction_constraints_theorem}.
\end{proof}

\end{document}